\documentclass[11pt,twoside]{article}
\usepackage{asp2010}

\resetcounters

\begin{document}

\title{A new look at the symbiotic star RW Hydrae}
\author{Magdalena Otulakowska-Hypka$^1$, Joanna Miko{\l}ajewska$^1$, and Patricia A. Whitelock$^{2,3}$
\affil{$^1$Nicolaus Copernicus Astronomical Center, Polish Academy of Sciences, Warsaw, Poland}
\affil{$^2$South African Astronomical Observatory, P.O. Box 9, 7935 Observatory, South Africa}
\affil{$^3$ Astronomy, Cosmology and Gravity Centre, Astronomy Department, University of Cape Town, 7701 Rondebosch, South Africa}}

\begin{abstract}
We present new estimates of the basic stellar parameters of the non-eruptive, eclipsing symbiotic system, RW~Hydrae. 
A set of photometric and spectroscopic data was used to model this object simultaneously from the light and radial velocity curves.
With new spectroscopic data we were able to improve previous results known from the literature and derive physical parameters of the system: 
 $q=4.2$, 
$M_1=0.8 \rm M_\odot$, 
$M_2=3.4 \rm M_\odot$,
$R_1=0.2 \rm R_\odot$,
$R_2=145 \rm R_\odot$,
$a=350 \rm R_\odot$, and 
$i=75^\circ$.

\end{abstract}

\section{Introduction}

Symbiotic stars are close binary systems with orbital periods of the order of years. In these objects an evolved red giant star is the secondary, and a hot companion (usually a white dwarf) is the accretor of material lost by the red giant. This accreted material powers symbiotic activity, including occasional eruptions and jets. 

One of the well known, non-eruptive symbiotic systems is RW~Hydrae (RW~Hya). This is an eclipsing object, thus it is relatively easy to derive its basic parameters. 
The spectroscopic orbit of RW~Hya is known from the literature: it has rather short orbital period, $\rm{P_{orb} = 370.2 \pm 0.9}$~days, and contains M2 giant and a hot white dwarf \citep{1995Kenyon}.
The near infrared (NIR) light curves show a modulation at half the orbital period \citep{2007Rutkowski} caused by the ellipticity of the red giant \citep{2002Mik}.
It is interesting that RW~Hya is metal-poor [M/H]$=-0.5\pm0.1$ and has a high proper motion.
This, together with the fact that it is located well out of the Galactic plane, at $l=314.9926, b=+36.4856$, suggests that it is in the halo (Miko{\l}ajewska et al. in preparation).

Our motivation for this study was the fact that we already had a number of photometric and spectroscopic observations of the red giant component of RW~Hya, which were published before \citep{1950Merrill, 1995Kenyon, 1996Schild, 2007Rutkowski}. 
Additionally, we used optical and UV spectroscopic data of HeII lines, which allowed us to trace also the orbital motion of the hot component. What is more, we wanted to check out one of the most recent tools for modelling eclipsing binaries, the \textit{PHOEBE} software. 

\section{Observations}

Our $JHKL$ photometry was obtained on the 0.75~m telescope at SAAO and is on the photometric system defined by \cite{1990Carter};
it has been discussed previously by \cite{2007Rutkowski}. 
The uncertainty of individual measurements is less than 0.03~mag in \textit{JHK} bands, and less than 0.05 mag in \textit{L}.

The spectroscopic data of RW~Hya come from several sources:
\begin{itemize}
\checklistitemize
\item HeII 4686 line -- from \cite{1950Merrill},
\item HeII 1640 line -- from IUE and HST (GHRS \& STIS) archives,
\item red giant data -- from \cite{1950Merrill}, \cite{1995Kenyon}, \cite{1996Schild}.
\end{itemize}

Light curves are shown in Fig.~\ref{fig-lc} and Fig.~\ref{fig-lc2}. Radial velocity curves are presented in Fig.~\ref{fig-rv} and Fig.~\ref{fig-rv2}. Different symbols are used to represent different data sources (with different precisions). Measurements from the IUE are the least accurate ones. In addition, we cannot exclude some offset between optical and UV data of the hot component.

\section{Analysis with \textit{PHOEBE}}

The \textit{PHOEBE} software is one of the recent tools for modelling eclipsing binary stars \citep{2005Prsa}. It is based on the Wilson-Devinney model \citep{1971Wilson}. 
The main feature of this code is the fact that it allows us to solve for all light curves simultaneously with the radial velocity data. 

\subsection{Initial assumptions}
\label{sec-IA}

From the literature we knew the effective temperature ($T_{eff}$) of the red giant from the spectral type M2 (3700~K), $T_{eff}$ of the hot component (30000~K) and metallicity $\rm{[M/H]} = -0.5 \pm 0.1$  (Miko{\l}ajewska et al. in preparation). 
From the spectroscopic orbit solution, which was based on red giant data, we could assume initial values for the  orbital period, velocity of the center of mass, superior conjunction, ranges of masses, mass ratio and eccentricity \citep{1950Merrill, 1995Kenyon, 1996Schild}. 

Additional assumptions:
\begin{enumerate}
\item The red giant is the only source of radiation in the NIR.
\item The safest option was to assume an unconstrained binary system in \textit{PHOEBE}.
\item The hot companion was set as the primary star, and the red giant as the secondary.
\item Limb darkening coefficients based on tables by \cite{1993vH} were used as implemented in \textit{PHOEBE}.
\item For both components we applied gravity darkening exponents and bolometric albedos for objects with convective envelopes 
($g_1 = 0.32$, $g_2 = 0.32$, $A_1 = 0.5$, and $A_2 = 0.5$). 
For the hot component we also obtained a solution for the radiative envelope 
($g_1 = 1.0$, $A_1 = 1.0$) but there was no difference between these two in the final result.

\end{enumerate}

\subsection{Analysis}

Our strategy for modelling with \textit{PHOEBE} was the following:
\begin{enumerate}
\item careful assumptions (see Sec.~\ref{sec-IA}),
\item iteration steps around the assumed values:
\begin{itemize}
\checklistitemize
\item radial velocity curves alone -- for solving system elements,
\item both, light curves and radial velocity curves -- to find physical parameters of components.
\end{itemize}
\end{enumerate}

At the beginning we had some difficulties with the modelling. Especially the spectroscopic data was hard to follow. 
Since previous studies of this object were based only on the red giant data, our initial assumptions could be incorrect.
Thus, we decided to set some constraints on the mass ratio from both radial velocity curves separately. 
The circular orbit fit to the radial velocity data of the red giant (RG) results in an orbital semiamplitude, $K_{RG}=8.7\pm0.3~\rm{km~s^{-1}}$, and the systemic velocity, $\gamma_{RG}=12.4\pm0.3~\rm{km~s^{-1}}$. 
While the fit to the radial velocities of the hot component~(HC) results in 
$K_{HC}=41.4\pm3.8~\rm{km~s^{-1}}$ and $\gamma_{HC}=5.6\pm3.6~\rm{km~s^{-1}}$. 
From these two orbital semiamplitudes we found $q={K_{HC}}/{K_{RG}}=4.8\pm0.6$, which was used as the restriction on the mass ratio in the first part of our analysis.
In the next section we present both results: with and without the constraints.

\section{Results}

\begin{figure}
\includegraphics[width=0.175\textwidth,angle=270]{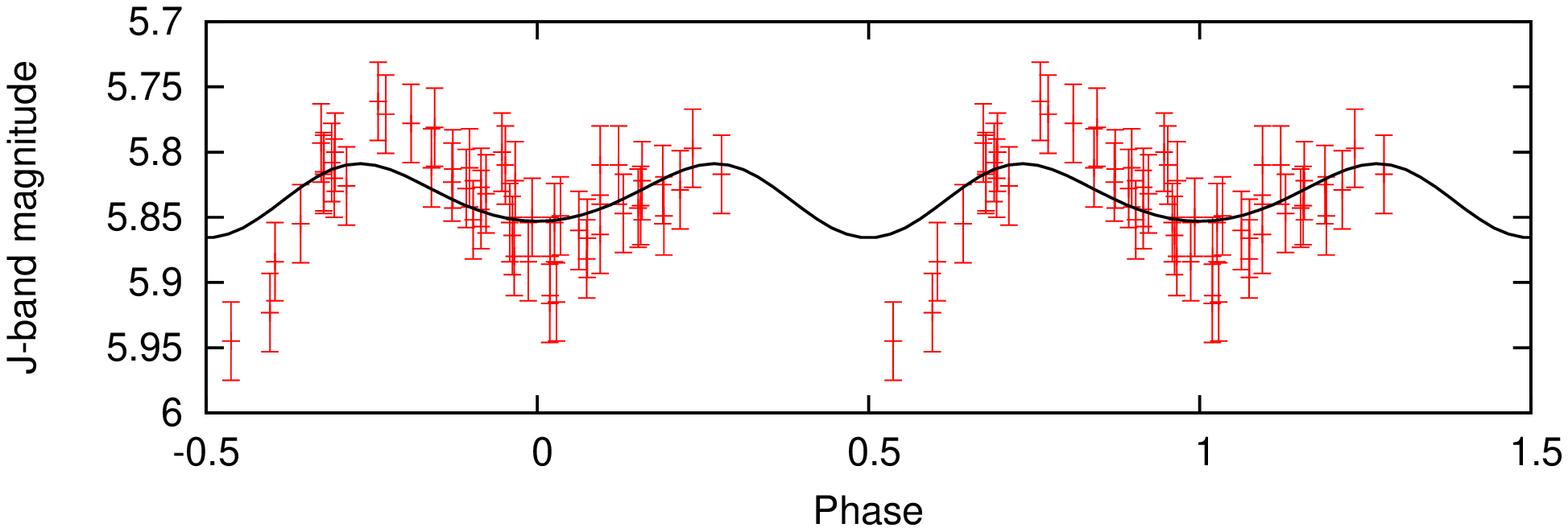}
\includegraphics[width=0.175\textwidth,angle=270]{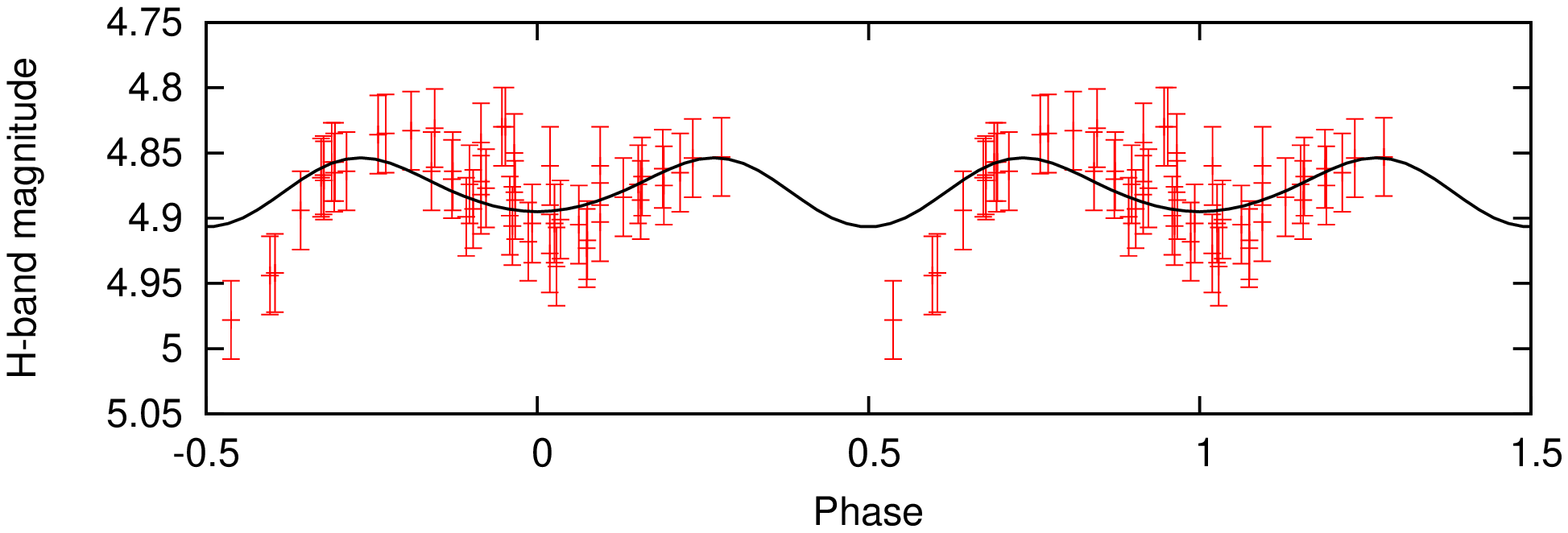}\\
\includegraphics[width=0.175\textwidth,angle=270]{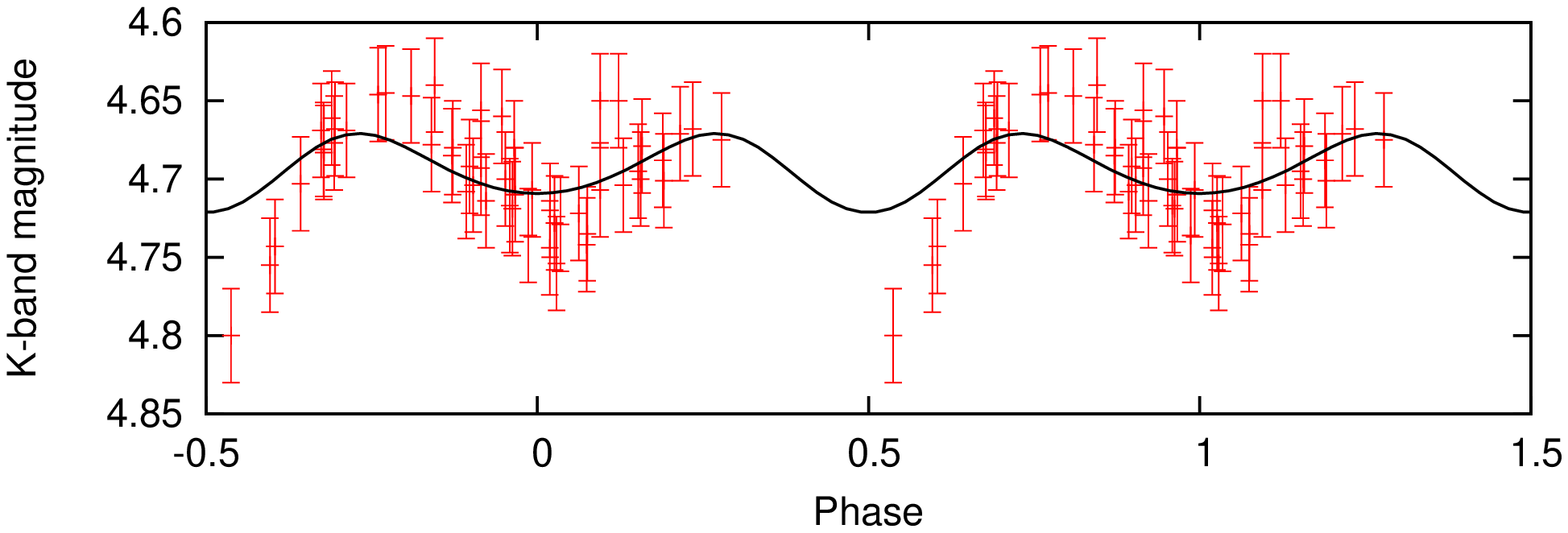}
\includegraphics[width=0.175\textwidth,angle=270]{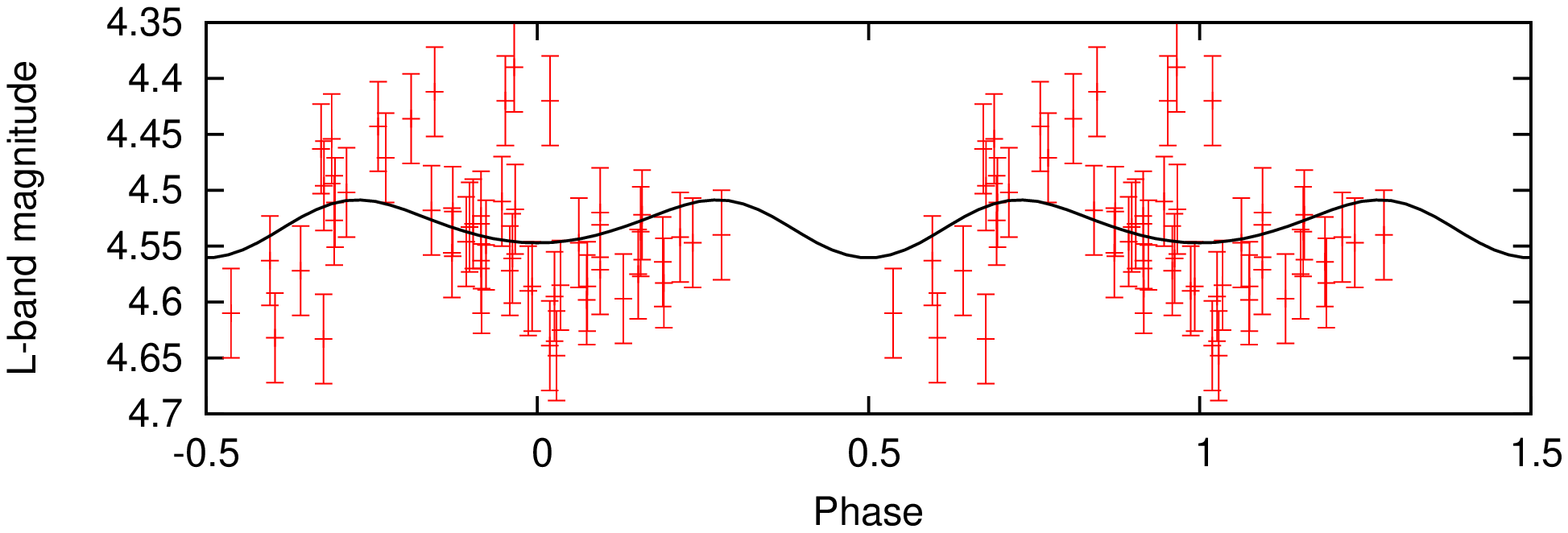}
\caption{Results for light curves with constraints.}
\label{fig-lc}
\end{figure}

\begin{figure}
\includegraphics[width=0.175\textwidth,angle=270]{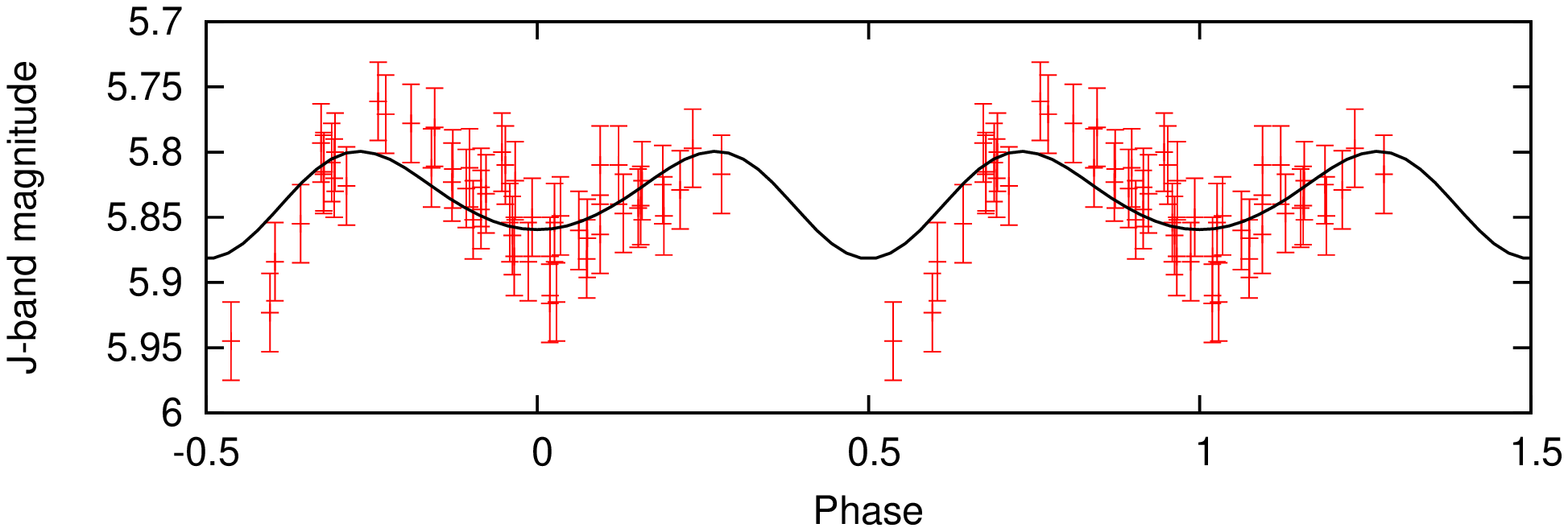}
\includegraphics[width=0.175\textwidth,angle=270]{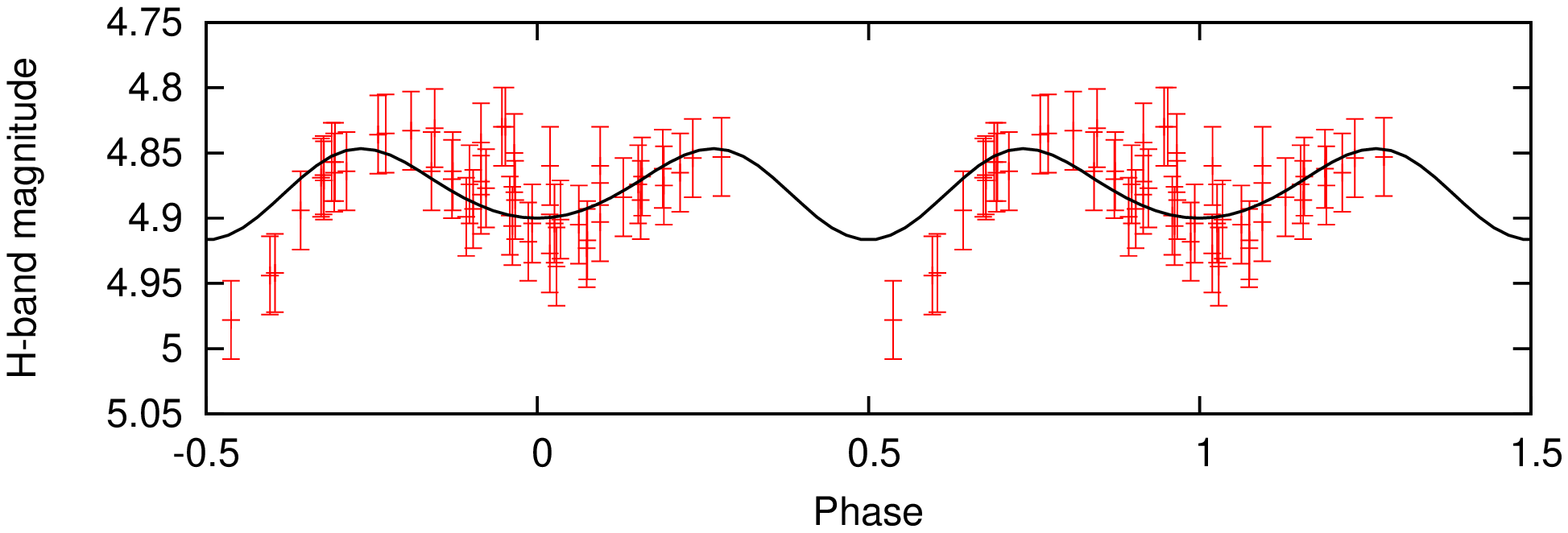}\\
\includegraphics[width=0.175\textwidth,angle=270]{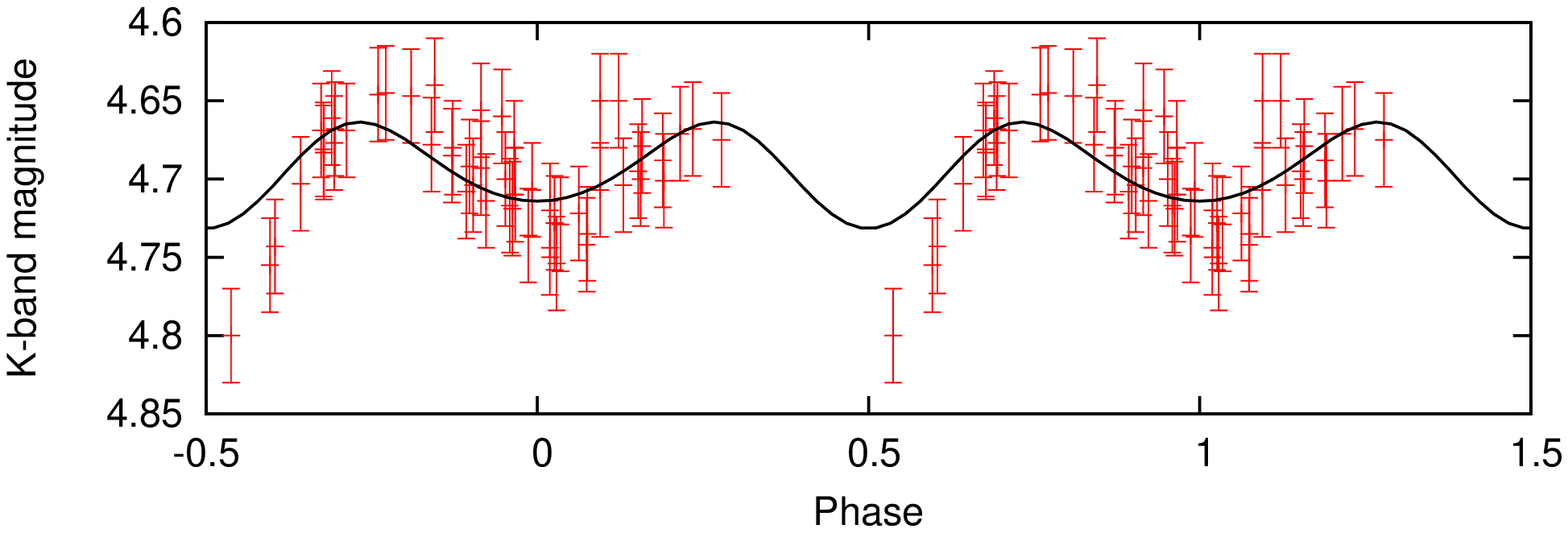}
\includegraphics[width=0.175\textwidth,angle=270]{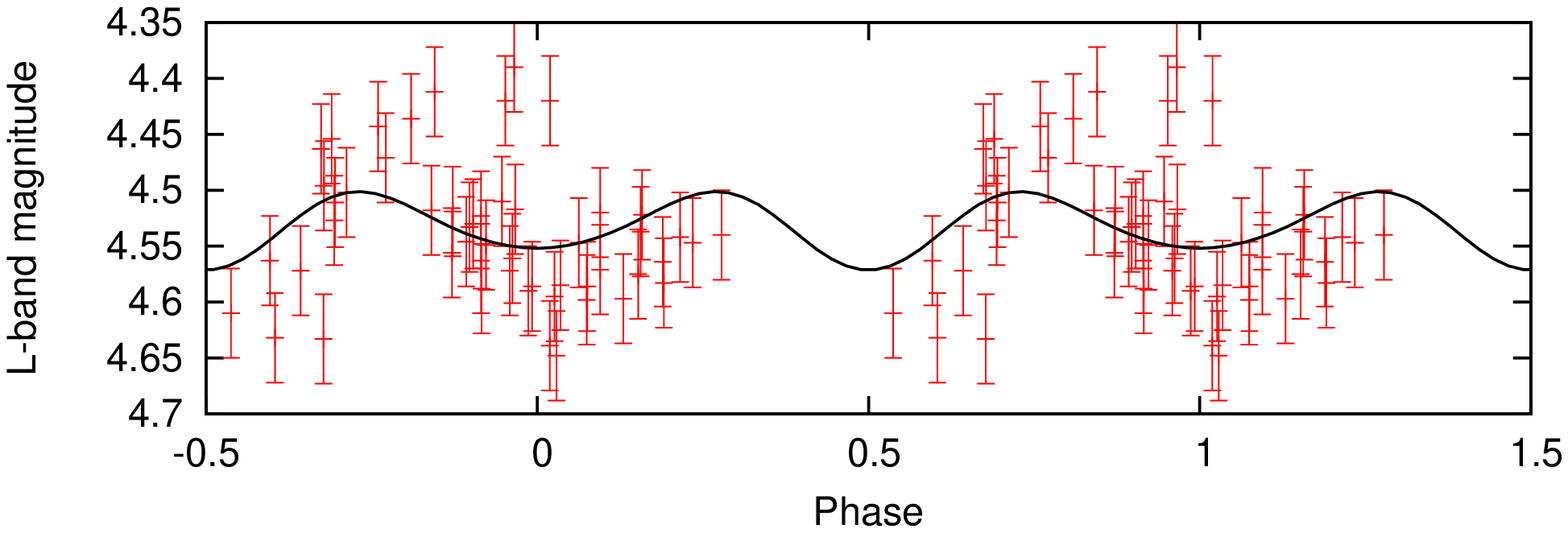}
\caption{Results for light curves without constraints.}
\label{fig-lc2}
\end{figure}

\begin{figure}
\includegraphics[width=0.4\textwidth,angle=270]{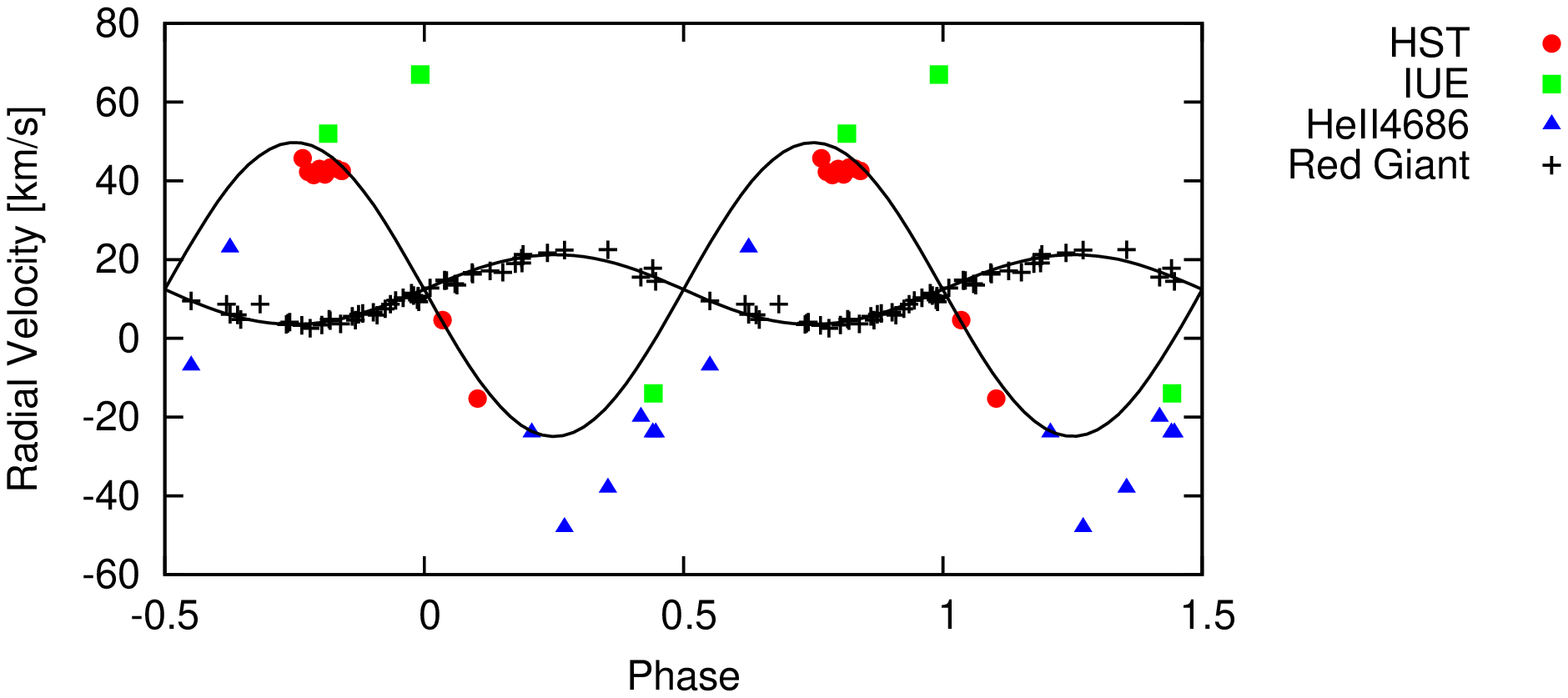}
\caption{Results for radial velocity curves with constraints.}
\label{fig-rv}
\end{figure}

\begin{figure}
\includegraphics[width=0.4\textwidth,angle=270]{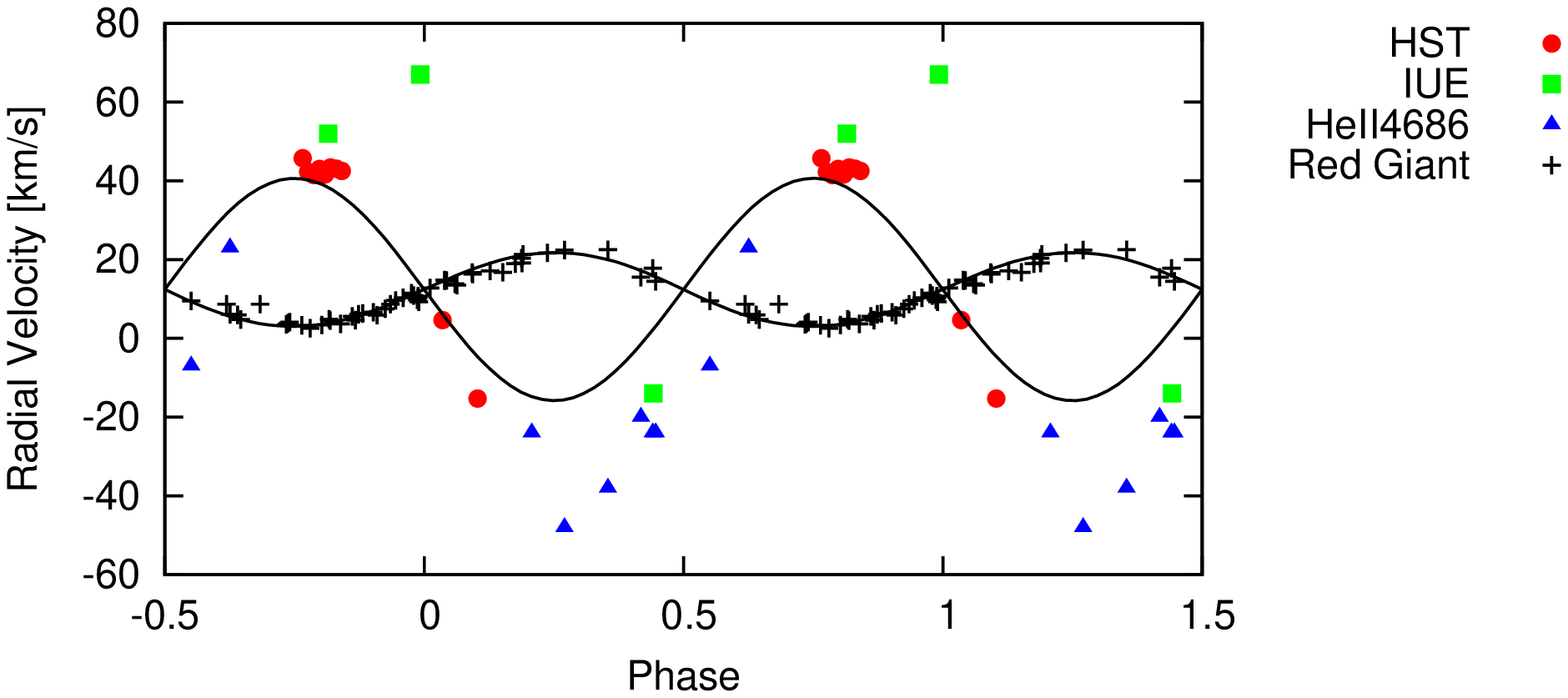}
\caption{Results for radial velocity curves without constraints.}
\label{fig-rv2}
\end{figure}

\begin{table}
\caption{Results for both cases: with and without the constraints}
\label{tab-res}
\begin{tabular}{c|c|c}
\hline 
& with constraints & without constraints  \\ 
\hline 
\textit{a} [$\rm R_\odot$] & 350 & 280 \\ 
\textit{i} [$^\circ$] & 75 & 80 \\ 
\textit{q} & 4.2 & 3 \\ 
$V_o$ [km/s]& 12.4 & 12.4 \\ 
$M_1$ [$\rm M_\odot$]& 0.8 & 0.5 \\ 
$M_2$ [$\rm M_\odot$]& 3.4 & 1.6 \\ 
$R_1$ [$\rm R_\odot$]& 0.24 & 0.16 \\ 
$R_2$ [$\rm R_\odot$]& 144.5 & 111.9 \\ 
$RL filling$ & 0.96 & 0.95 \\ 
\hline
assumed: & & \\
\hline
$P_{orb}$ [d] & 370.2 & 370.2 \\
$T_{eff}^1$ [K] & 30000 & 30000 \\
$T_{eff}^2$ [K] & 3700 & 3700 \\
\hline 
\end{tabular} 
\end{table}

In Fig.~\ref{fig-lc} and \ref{fig-lc2} we show both outcomes for the photometric data. Results without constraints seem to represent the observations better. 
This is primarily because the unconstrained solution results in an inclination that is five degrees larger than the constrained solution.

Fig.~\ref{fig-rv} and \ref{fig-rv2} illustrate solutions for the spectroscopic data. Here the situation is
reversed: the outcome with constraints seems to be better, and we favour this result. 
The values obtained of the main parameters are presented in Table~\ref{tab-res} for both stars.

To find a unique solution for RW~Hya, we need more spectroscopic data. Fortunately, we are about to collect such observations in the near future. Then, we hope to improve this modelling. 

\acknowledgements 
MOH acknowledges support provided by the Organizers which allowed her to participate in the conference.
The project was supported by the Polish National Science Center grants awarded by decision number DEC-2011/03/N/ST9/03289 and 
DEC-2011/01/B/ST9/06145.
PAW acknowledges a grant from the National Research Foundation (NRF) of South Africa.

\bibliography{magdaot-talk}

\end{document}